\documentclass{article}
\usepackage{shadethm}
\newshadetheorem{definition}{Definition}
\newshadetheorem{observation}{Observation}
\usepackage{latexsym}
\usepackage{amsmath}
\usepackage{amssymb}
\usepackage{graphicx}
\usepackage[pdftex,
            pagebackref=true,
            colorlinks=true,
            linkcolor=blue
           ]{hyperref}
\usepackage{float}
\usepackage{alltt}

\usepackage{enumitem}
\begin{document}

\setlist[description]{font=\normalfont\normalfont\space}
\title{A few reflections on the quality of emergence in complex collective systems}
\author{Vincenzo De~Florio\\
  \multicolumn{1}{p{.8\textwidth}}{\centering\emph{
MOSAIC research group\\
       iMinds research institute \& University of Antwerp\\
       Middelheimlaan 1, 2020 Antwerp, Belgium\\
\makeatletter
       \textsf{vincenzo.deflorio@gmail.com}
\makeatother
  }}}

\date{\today}
\maketitle
\begin{abstract}
A number of elements towards a classification of the
quality of emergence in emergent collective systems are provided. By using those
elements, several classes of emergent systems are exemplified, ranging from simple aggregations of
simple parts up to complex organizations of complex collective systems.
In so doing, the factors likely to play a 
a significant role in the persistence of emergence and its opposite are highlighted.
From this, new elements for discussion are identified also considering elements
from the philosophy of Leibniz.
\end{abstract}

%%%%%%%%%%%%%%%%%%%%%%
\section{Introduction}
%%%%%%%%%%%%%%%%%%%%%%

Emergence: that ``magic'' that produces something new and more
from the interactions of a set of parts---sometimes something
totally unexpected and even bewildering. Something that manifests itself with a ``strength'',
or an energy, or a \emph{meaning}, that is
way greater than the one exhibited by the individual parts that make the whole up.

Emergence as well as the factors responsible for its manifestation
are the main characters in this work. My focus shall be quite general and
will try to characterize the emergence that manifests itself
from collections of simple objects up to complex organizations
of complex systems.
Examples shall be considered among artificial, natural, and social systems.
Far from an exhaustive treatise, this may be considered as a first step
towards a characterization of those factors that sustain the manifestation of emergence
in complex collective systems when subjected to change.
The word ``change'' shall assume meanings that are specific
of the considered level, or ``scale,'' of the system at hand:
\begin{itemize}
\item In the case of simple objects and systems, change shall refer to, e.g.,
      the consequences of wear-out, defects, physical failures, or
      operational conditions beyond those expected or foreseen is some specification
      document.
\item In the case of more complex collective systems, change shall include
      the effect of forces that disrupt or strengthen the cohesion of the social system.
\end{itemize}
I will use the term ``\emph{quality of emergence}'' (QoE) to refer to the cumulative effect
of the above factors and changes.

%Does it make sense to talk of a resilience of emergence? 1) Identity of emergence, and 2) Its persistence
%in the face of change?
%
%How do we sustain that property?

%First of all, we need to highlight that this is something that
%has been known for a long time. Aristotle already discussed it in his Metaphysics\cite{AristotleMetaphysics}.
%Metaphysics---something that goes beyond what is physical, material, as the word already seems to suggest.

In what follows I first present, in Sect.~\ref{s:prelim}, three elements useful for
the characterization of the QoE in complex collective systems.
Those elements are then used in Sect.~\ref{s:elements} to define several classes
of emergent systems.
%Leibniz?
Some remarks on related concepts---resilience and failure semantics---are
then given in Sect.~\ref{s:dis}.
Preliminary lessons learned are finally summarized in Sect.~\ref{s:end}.

%%%%%%%%%%%%%%%%%%%%%%%%%%%%%%%%%%%%%%%
\section{Preliminaries}\label{s:prelim}
%%%%%%%%%%%%%%%%%%%%%%%%%%%%%%%%%%%%%%%
It is already emergence something that appears to ``spring'' from the simple interactions of
simple interconnected objects. Let us make use of an example. If you consider a car and its parts,
we see that among all of the possible combinations of the latter there are some that allow the union of the parts
to become something else, something \emph{greater}; something that makes the new
element respond to a new and specific definition: in this case, the definition of car.

%Let us take note of what we had in this example: a \emph{combination\/} of parts. 
%Such concept plays an important role in our discussion, and I will return to it later.

Here is a first, simple example of emergence: a tool that is produced by the combination of a set of components.
We observe how in this case, from a systemic point of view, the compound differs only slightly from the components.
Both components and compound are ``simple'' systems. Cogs made of cogs.
Futhermore we observe how even the ``glue'' that holds together the components and,
in fact, \emph{realizes\/}  the tool, is in this case very much simple:
it is made of physical connections based on the material properties of the components;
on their shapes; and on auxiliary components the purpose of which is to ensure the persistence
of the identity of the compound; to ``keep it together'', at least for a certain period of time
and under certain operating conditions.

The name of this ``glue'' is \textbf{organization}. Emergency is therefore a property of a
compound formed from components organized into a \emph{whole}. Organization is what holds together
and puts in relation the components with one another.
My stance here is that these three aspects---the systemic characteristics of
components, of the compound, and of the ``glue''---constitute three of the key elements
to characterize an emerging system and its quality.

A first aim of this paper is to lay the foundation to a future
classification of ``emergent systems'' based on the just mentioned triplet.

Two methodological assumptions are introduced in the rest of this section.

%-----------------------------------------------------------%
\subsection{Methodological assumptions and preliminary steps}
%-----------------------------------------------------------%

Components, compound, and organization are three elements that I selected to operate
a classification of emergent systems. In order to distinguish among compound and component systems,
in what follows I propose to use a general systems classification such as the one introduced
in~\cite{RWB43} or~\cite{Bou56}, or their augmentations~\cite{DF15b}. By means of one such
set of classes it is possible to introduce the following definitions.

\begin{definition}[systemic quality]
In what follows I shall refer to the systemic quality of a component or a compound as to one
of the following classes:\label{s:metric}

\begin{description}
\item[\mbox{P}:]
systems only capable of passive behavior; in other words, objects.
\item[\mbox{$\neg$T}:]
 systems capable of purposeful, non-teleological behavior---so-called servo-mechanisms.
\item[\mbox{T}:] 
teleological (that is, reactive) systems. These are systems that embed a feedback loop.
\item[\mbox{T}$^{+}$:] more-than-teleological systems. This class includes,
among others, extrapolative, learning, autonomic, adaptive, and antifragile systems~\cite{Taleb12,DF14a,Jones2014870}.
\end{description}
\label{d:sq}
\end{definition}

\begin{definition}[total order on the set of systemic qualities]
A total order on the set of systemic classes mentioned in Definition~\ref{d:sq}
is introduced as follows:

\[ \mbox{P} \,\prec_\sigma\, \mbox{$\neg$T} \,\prec_\sigma\, \mbox{T} \,\prec_\sigma\, \mbox{T}^{+}. \]
\label{d:tosq}
\end{definition}

For any two systems $a$ and $b$, if $a \prec_\sigma b$ we shall say that $a$ is \emph{less evolved\/} than $b$.
To refer to the systemic quality of any system $a$ we shall use notation $\sigma(a)$.

What just done for systems is now done for organizations.

\begin{definition}[Organizational quality]
I shall refer to the organizational quality of a set of components realizing
a compound as to one of the following classes:

\begin{description}
\item{$+$\hskip5pt} Juxtaposition. This is not an organization \emph{sensu proprio}; rather, it
is a simple grouping of elements. Emergence is in this case a consequence of the additive property
of a physical quantity.
\item{$\Xi$\hskip7pt} Strictly hierarchical organizations,
characterized by top-down flow of control and bottom-up flow of feedbacks (autocracy).
Several traditional human-based organizations
are based on this class.
\item{$\Xi^{+}$} Nested compositional hierarchies. 
This type of organization is based on so-called ``principle of increasing inclusiveness'', in which 
``entities at one level are composed of parts
at lower levels and are themselves nested within more extensive entities''~\cite{HT:TE14a}.
Widely adopted in nature, class $\Xi^{+}$ makes a systematic use of modularity, one of the key factors
leading to the emergence of complexity---what is defined in~\cite{WaAl1996} as
``the evolution of evolvability.''
\item{$\Xi^{\times}$} This is the class of what I call ``metarchies'': more-than-hierarchical organizations.
This includes, among others, heterarchies~\cite{Stark} and exception-based
hierarchies~\cite{DF15a}, such as  sociocracy~\cite{BuEn12} and
socio-fractal organizations~\cite{DF13a,DFSB13a,DeFPa15a,DeFPa15b}.
\end{description}
\label{d:oq}
\end{definition}

As it has been done for the systemic quality,
it is now possible to introduce a total order for organizations.

\begin{definition}[total order on the set of organizational qualities]
I define a total order between the classes introduced in Definition~\ref{d:oq}
as follows:

\[ + \,\prec_o\, \Xi \,\prec_o\, \Xi^{+} \,\prec_o\, \Xi^{\times}. \]
\end{definition}

If $a \prec_o b$ I shall say that $a$ is less evolved than $b$.
In what follows I shall use
symbol $o(a)$
to refer to the organizational quality of any organization $a$.

To simplify the notation, when this can be done without introducing ambiguity, I
 will make use of symbol ``$\prec$'' to refer to either of 
$\prec_\sigma$ and $\prec_o$.

In what follows, I shall refer to the emergent systems as either Whole or Compound
and to its elements as either Parts or Components.

%%%%%%%%%%%%%%%%%%%%%%%%%%%%%%%%%%%%%%%%%%%%%%%%%%%%%%%%%%%%%%%%%%%%%%%%%%%%%%%%%%%%
\section{Elements of a classification of emergent systems}\label{s:elements}
%%%%%%%%%%%%%%%%%%%%%%%%%%%%%%%%%%%%%%%%%%%%%%%%%%%%%%%%%%%%%%%%%%%%%%%%%%%%%%%%%%%%
Here I provide some elements of a classification of emergent systems.

One of the lower levels, if not the lowest, is the one of systems operating
as LED lamps. Those are systems in which
\begin{itemize}
\item \emph{simple components\/} (more precisely, systems whose behavioral class is at most \mbox{$\neg$T})\ldots
\item \emph{\ldots implement a system as simple as their components\/} (again, at most \mbox{$\neg$T},
although characterized by non-trivial emergent properties)\ldots
\item \emph{\ldots by means of a simple organization\/} (``simple'' meaning here
``at most of class $\Xi$'').
\end{itemize}

A LED lamp is in fact composed from simple\ldots LED's, which, however,
constitute a more powerful light source. Interestingly enough, such 
light source is characterized by \emph{additional\/} properties---properties that are
not present in the constituent elements. For instance, a LED lamp is characterized
by low power consumption.

But there also more ``subtle'' and I would say more
intriguing properties that in this case characterize and \emph{distinguish\/} the compound
from the components---the Whole from its Parts.
One such property is \emph{graceful degradation}.
While an ordinary lamp has a ``binary'' failure semantics (it is either ``working'' or not at all),
a LED lamp has what I would refer to as a \emph{fuzzy\/} failure semantics. In fact,
there is an ample spectrum of transition states in between ``working'' and ``not working''!
I believe it is interesting to highligh that in this case
the ``glue'', namely the organization of the LEDs in an LED lamp, is perhaps the simplest possible.
The ``cohesion'' that allows the Whole to emerge from its Parts is in fact based on
simple physical laws. The light expressed by the Whole is produced through additive synthesis of 
the lights emitted by the Parts----which, among other things, makes it possible
to easily vary the emerging color of the lamp~\cite{Schub06}. By referring to the metric proposed
in Sect.~\ref{s:metric}, organization is in this case of class ``$+$''.

We can conclude that a LED lamp provides us with a first ``class'' of emergent systems:
\begin{equation}
(\mbox{$\neg$T}, \mbox{$\neg$T}, +).\label{c:led}
\end{equation}
Triplet~\eqref{c:led} is
representative of systems in which both compound and components are characterized by
at most pourposefil, non-teleological behaviors~\cite{RWB43}
and the organization is of the additive class $+$.
As already pointed out, more than of an organization it may make more sense in this case to
talk of a ``grouping''---a combination of the Parts.
%LEIBNIZ%
It is worth remarking that  the emergent properties gradually appear
with the juxtaposition of the components, and gradually disappear with their removal or their disintegration.

\begin{definition}
Emergent systems characterized by the triplet~\eqref{c:led} will be called in the following LED systems.
\end{definition}

In what follows I shall not engage in a comprehensive classification, whose attempt
I aim to address in subsequent works. Aim of this paper is 
to pave the way for such an endeavour by defining some classes of emergent systems
of particular interest---as well as to reflect about possible relationships between
those classes and emergence ``meta-properties'' (namely, properties related to the
property of emergence.)

\subsection{Hierarchies with Components less advanced than the Compound}
Here I will discuss the family of emergent systems that are characterized by the adoption of a
hierarchy (thus an organization of either $\Xi$ or $\Xi^{+}$ class) and with $o(\mbox{Parts}) \prec o(\mbox{Whole})$.
Again, this is not meant as an exhaustive treatise---rather, it is an exemplification
of several of the members of a family of emergent systems.

\subsubsection{Organs}
I shall call ``organs'' those emergent systems that are made of
passive-behaviored components (``objects'', or ``cogs'')
that are assembled into a purposeful, non-teleological behaviored Whole---for example, a servomechanism.
I shall use the following notation:
\begin{equation}
(\hbox{P}, \mbox{$\neg$T}, \,\Xi),\label{c:organ}
\end{equation}
``Assembled'' here stands for a rigid and immutable hierarchy (namely, class $\Xi$).

\begin{definition}
Emergent systems characterized by the~\eqref{c:organ} triplet will be called in the following 
as \textbf{organs}.
\label{d:organ}
\end{definition}

In the case of organs, the identity of the system (and thus the persistence of its ``quality'') is
purely related to the mechanical cohesion of the parts within the whole. In other words, there are no
``centrifugal'' forces~\cite{Dominici11,Dominici13,Dominici14}
leading the individual Parts to disrupt cohesion, because the parties are mere objects.
Persistence of identity and emergence is affected by wear-out, defective parts, defective assemblage, etc.

\subsubsection{Systems of organs}
In this class of emergent systems, organs are assembled to compose an autonomous system
capable of teleological / reactive behaviors~\cite{RWB43}.
In this case I shall use notation

\begin{equation}
(\mbox{$\neg$T}, \hbox{T}, \Xi).\label{c:teleo}
\end{equation}

\begin{definition}
Non-teleological Parts hierarchically united to form a teleological system, as expressed by the \eqref{c:teleo}
triplet, shall be called in what follows a \textbf{system of organs}.
\label{d:teleo}
\end{definition}

The autonomic nervous system of animals is an example of a system of organs.
Persistence of resilience is usually a result of
wear-out, malfunctioning, defective organs, or external conditions.

\subsubsection{Organisms}
Systems of organs,  organized into autonomous systems characterized by autonomicity, proactivity,
and other advanced capabilities
(in some cases sentience, self-awareness, and antifragility~\cite{Jones2014870,DF15b}),
are called in what follows \textbf{organisms}.
More formally, an organism is characterized by the following triplet:
\begin{equation}
(\hbox{T}, \hbox{T}^{+}, \Xi^{+})\label{c:organism}
\end{equation}
and is defined as follows:

\begin{definition}
Teleologically behaviored Parts producing a more-than-teleologic Whole
and organized as nested compositional hierarchies as expressed by the triplet~\eqref{d:organism}
shall be called in what follows \textbf{organisms}.
\label{d:organism}
\end{definition}
A fractal organization of systems of organs, as exemplified by the human body, \emph{realizes\/}
an organism.

Organisms may be characterized by a ``body''---for instance, a physical, a legal, or a social ``body''. Depending
on the nature of its body, an organism may be more or less sensitive
to disgregative forces such as wear-out, aging, external threats, parasitivism, and extreme ``individualism.''

%AGRIPPA%

\subsubsection{Societies}
In what follows, an organization of organisms shall be called a society. More formally:
\begin{definition}
A society is an emergent system characterized by the following triplet:
\begin{equation}
(\hbox{T}^{+}, \hbox{T}, \,\,\xi),\label{c:society}
\end{equation}
$\xi$ being \emph{any type\/} of organization.
\label{d:society}
\end{definition}
As apparent from Definition~\ref{d:society}, societies are a peculiar case
of emergent systems, in that they are an organization of highly evolved Parts
that produce a Whole that is less evolved than its Parts. What is also remarkable is that said
organization may vary across the whole range of organizational classes.

I shall now focus my attention on four sub-classes of societies, defined as follows.

\begin{definition}[Parasitic society]
An organization of organisms (i.e., a society) that provides returns that are
beneficial to only some of the involved organisms shall be called
a \textbf{parasitic society}.
\end{definition}
Sentinel species~\cite{dS99+} are an example of organisms \emph{employed\/} for
a purpose defined by other organisms. Canaries for instances have been used
for a long time to alert miners of the presence of 
of dangerous concentrations of
toxic substances---e.g., carbon monoxide and dioxide, or methane~\cite{DF15b}.
In parasitic societies,
persistence of emergence is guaranteed by the impossibility for the Parts to
leave the Whole. Other examples that come to mind are that of
social systems including so-called slaves,
as it was the case, e.g., in ancient Egypt, or social systems including
a class deprived of most of the civil rights, as it was for instance for the \emph{Tiers \'Etat},
the weakest of the estates in the organization of the French state
before the Revolution.

\begin{definition}[Ecosystem]
A society that sustains returns that are
mutually satisfactory for the Parts and the Whole, and that is beneficial to all parties involved---without
explicitly privileging some of the parties involved---shall be called
in the following an \textbf{ecosystem}.
\end{definition}
An exemplary ecosystem is a beehive~\cite{Mae1910}. In this case
\emph{a mutualistic relationship exists between Parts and Whole}.
This is similar to what happens in nature across the scales of natural systems.
In fact this phenomenon takes place even between societies---the natural kingdom of
animalia and that of plantae being a well-known example.

The organisms of an ecosystem are able to establish a \emph{harmony\/} of sorts,
recognizing the role that the Whole plays for the Parts and vice-versa.
Persistence of emergence ``emerges naturally'' because of said sustained harmony.

An interesting paradox is that said harmony is stronger when the organisms that
constitute the society have not developed a strong sense of individuality---thus their persona
can more easily blend with the persona of the Whole~\cite{DF14d}.
Human societies are a classic case in which disharmony between Parts and Wholes may manifest
itself. The resulting ``tragedies of the common''~\cite{Hardin68} have been impacting
severely on the sustainability of our species.

\begin{definition}[Factory System]
A third sub-class of societies is a \textbf{factory system}, namely
one that provides unbalanced returns that,
although being mutually beneficial, are biased towards
a subset of the Parts.
\end{definition}

The Industrial Revolution and subsequent times produced many extreme cases of factory systems---ironically
rendered by Charles Chaplin in his renowned ``Modern Times'' as a human organism caged into a production
mechanism.

Remarkable traits of a factory system are the minimal or absent
identification of the components with the compound. The Parts typically do not blend into the Whole
while the Whole often does not recognize or value the individuality of the Parts, which are regarded in the
same way as interchangeable parts in a manufacturing system. Cohesion and persistence
of emergence are a consequence of environmental
conditions that motivate the parts to accept their role of components despite the unbalanced returns.

\begin{definition}[Defense System]
A fourth sub-class of societies is a \textbf{defense system}, namely a society
of organisms whose major sought after return is survivability of the Whole and the Parts
in the face of an external threat.
\end{definition}
The external threat (the ``foe'') personifies the reason for cohesion and reduces
centrifugal forces despite the adoption of rigid forms of hierarchy (typically, $\Xi$).
Persistence of emergence is also strengthened by specific internal regulation (in the
case of armed forces, this is called ``military justice'').

%%%%%%%%%%%%%%%%%%%%%%%%%%%%%%%%%%%%%%%%%%%%%%%%%%%%%%%
\section{Resilience and failure semantics}\label{s:dis}
%%%%%%%%%%%%%%%%%%%%%%%%%%%%%%%%%%%%%%%%%%%%%%%%%%%%%%%
\subsection{Resilience as a dynamic property}
In Sect.~\ref{s:prelim} I briefly hinted at a crucial aspect that has not been further developed
in the present article. I refer to resilience, namely
a compound's persistence of identity~\cite{DF15d,DF15b}.
There is clearly a strong link between QoE and resilience, as a system's identity
prescribes and details the emergence of a number of expected properties, traits, or
characteristics.

It is important to realize that, as QoE, also resilience is a \emph{dynamic\/} property:
a compound retains its characteristics ``for a certain period of time
and under certain operating conditions''.
Resilience is thus a trait that may appear,
be sustained for a while, and then disappear.
As a consequence, rather than considering QoE as an absolute
property, we should consider a reference QoE and match it
continuously with an actual, observed, QoE. The already mentioned example
of a car applies here too: the concept of car entails a number of
reference properties---for instance, the car should move as we expect it to do;
should be controllable by means of an agreed upon, standard interface; and so on.
When the observed properties of a car drift away from those reference properties,
the car looses its identity.
In other words, in order to be a car, a system should
``behave'' like a car. When the observed behaviors do not match anymore the expected ones,
the car is no more resilient.
The approach I followed
in~\cite{DF14a,DBLP:journals/corr/FlorioP15} is to measure resilience by considering
a drift from the reference quality---in fact, a distance.
Such drift, or distance, is also a dynamic system---thus a property that varies in time.
In the cited papers I highlighted how different resilience classes may be associated to
different ways for the drift to vary in time. A more fine-grained classification
of QoE should also consider the resilience class exhibit by the system under consideration.

\subsection{Resilience as a compound's property}
%``I believe these three elements---the systemic characteristics of
%components, of the compound, and also of the ``glue''---are three of the elements
%that characterize an emerging system and its quality'':
A way to model a compound's resilience is possibly given by considering
%ment in the emergence of resilience is the ``health'' or ``strength''
%of what we called in Sect.~\ref{s:prelim} as the glue that binds together the components into the compounds.
the \emph{energy\/} that binds together the components into the compound and allows them to
fulfill the roles prescribed by the organization. Said energy is finite. When it
goes below a given threshold, a component and possibly its compound lose their identity.
In order to prolongue the life of the compound, certain components may be replaced---as
an old tire is replaced with a new one, for instance. When such process can be done
autonomously, then the system is said to self-repair.

In some cases this replacement is difficult or impossible to achieve. We can
replace a ``heart'' though we can not replace a ``brain''. In such a case, the
system dis-integrates---it ``dies''.
A compound characterized by a certain aggregation quality breaks down into
its components, thereby loosing the identity-of-the-compound.
Like a castle of cards that fall down, the compound that once stood and now stands no more
becomes again ``simply a pack of cards''.
The ancient ones reflected on this very phenomenon and associated it with \emph{sadness}.
Asclepius's Lament in the Corpus Hermeticum~\cite{CorpusHermeticum} stands for the sorrow
associated with this loss of emergence\footnote{%
``This All, which is a good thing, the best that can be seen in the past, the present and the future, will be in danger of perishing; men will esteem it a burden; and then they will despise and no longer cherish
\emph{this whole of the universe}, incomparable work of God, glorious construction,
good creation made up of an infinite diversity of forms, instrument of the will of God who,
without envy, pours forth his favour on all his work, in which is assembled in one whole,
in a harmonious diversity, all that can be seen that is worthy of reverence,
praise and love.''~\cite{Coppens}.%
%http://www.philipcoppens.com/thelament.html
}---especially when induced by external forces;
with ``divinity'' leaving the living statues of Horus and ``returning from Earth to Heaven''~\cite{Coppens}.
Said ``divinity'', which Giordano Bruno calls ``profound magic'', is indeed the miracle
of the persistence of identity~\cite{BrunoSpaccio}.

\subsection{A note on failure semantics}
In Sect.~\ref{s:elements} we briefly mentioned the role of failure semantics---the way
a system loses its resilience and stops manifesting its emerging properties.
As we exemplified already, systems with the same purpose may fail very differently.
This change of failure semantics is what we observe when moving for instance
from the traditional concept
of car to that of drive-by-wire car: as already mentioned, although in both cases
the purpose is the same, not so are the organizations of the parts across the scales
of those systems. The fact that the purpose is the same should not trick us into believing
that both systems shall fail the same way. A mechanical car often exhibits graceful degradation.
Moreover,
if an electronic car bumps lightly into a wall, the identity of that car may have suffered
in a more subtle, less evident way than the identity of a traditional car.
In other words, electronic cars are inherently more ``fragile'' than traditional ones.

%%%%%%%%%%%%%%%%%%%%%%%%%%%%%%%%%%
\section{Conclusions}\label{s:end}
%%%%%%%%%%%%%%%%%%%%%%%%%%%%%%%%%%
In this paper I provided a number of elements towards a classification of the quality of emergence
in emergent collective systems. A number of classes of emergent systems has been
exemplified, ranging from simple aggregations of simple parts up to complex organizations
of complex collective systems. In each class we highlighted those factors that appear to play a
significant role in the persistence of emergence or its opposite. A lesson learned while
collecting the above results is that apart from classic causes---such as
deterioration or wear-out; design faults; defective or non-optimal components, etc.---a significant
role in the sustaining of emergence is \emph{harmony\/} between the Parts and the Whole.
Lack of said harmony translates in fact into disruptive forces that
minimize identification of the Parts with the Whole
and therefore tend to repel the former
from the latter.
My conjecture here is that indicators of disharmony may be defined by cosidering
possible mismatches between $\sigma(\hbox{Components})$ and $\sigma(\hbox{Compound})$,
as well as mismatches between $\sigma(\hbox{Components})$ and $o(\hbox{Organization})$.

From the above reasoning, a new element to the present discussion naturally \emph{emerges}:
a fourth element to the triplet (compound, components, organization) 
may be \emph{sustainability}, which I interpret here as
a collective system's propensity towards the realization of mutualistic relationships,
strengthening identification, and other ``centripetal'' forces.

A final remark brings me back to my first example, when I considered 
a car and its parts and I mentioned
that among all of the possible combinations of the parts there are some that allow the compound
to become something ``else''. I think it is important to remark here how
emergence indicates the birth of a new concept, of a new and unique ``substance''
responding to a new \emph{definition}---in Aristotelian and
Leibnitian terms~\cite{DBLP:journals/corr/Florio14d,DF14c,Burek04}.
Moreover, ``substance'' is the materialization / implementation / \emph{real}ization
of an abstract concept---what Leibniz calls a \emph{monad}; thus, it is
a ``physical'' rendition of a purely conceptual idea. It is again Leibniz that
suggests that QoE may be better assessed by considering an intrinsic,
an extrinsic, and a ``social'' element:
\begin{description}
\item[Intrinsic element:] The ``design choices'' of which elements to combine and how. This, in a sense,
is the ``abstract code'' of the system being considered.
\item[Extrinsic element:] The ``implementation choices'' of which ``physical'' parts; organs; organisms; and societies
 to employ in order to produce a material instance of the abstract code.
\item[Social element:] The ``compossibility'' (compatibility) of the material instance when
integrated in a society of other substances~\cite{leibniz2006shorter}; a concept that
anticipates Darwinian fitness and evolution.
\end{description}

%interchangeable parts

%nuovo elemento: ritorno ``economico'': sopravvivenza, benessere, servizi...

%calmiere delle forze disgregatrici,
%focolaio delle forze rafforzatrici

%All the above elements are missing in the current treatise and shall be the matter of future elaboration.
A preliminary discussion of the above elements is available in the draft paper~\cite{DF16L}
and shall be further elaborated in future contributions.

%In the exemplified case of a beehive,
%the organisms, although
%highly evolved, do not have a strong individuality---thus their persona can easily blend with the
%persona of the Whole~\cite{DF14d}.
%A remarkable aspect is that
%\begin{center}
%a mutualistic relationship exists between Parts and Whole.
%\end{center}
%In other words, parts and Whole constitute an ecosystem.

%[Remember: Another element will be the resilience of identity; cf. Leibniz. Match intrinsic and extrinsic. New "combination"; but with constraints (see compossibility)]
%[Then the "code"]

\bibliographystyle{plain}
\bibliography{/refs/thesis}
\end{document}